\documentclass[twocolumn,showpacs,preprintnumbers,
amsmath,amssymb]{revtex4}
          \usepackage{graphicx}
          
\begin{document}

          \title{"Artificial" superconductors. Superconducting phases in the $\textrm{Mg}_{x}\textrm{WO}_{3}$
          nanocomposite $(x = 0.037;\, 0.125$ - $T_{cx} = 140;\, 280$\!K ).}
          \author{V. N. Bogomolov}
          \affiliation{A. F. Ioffe Physical \& Technical Institute,\\
         Russian Academy of Science,\\
         194021 St. Petersburg, Russia}
         \email{V.N.Bogomolov@inbox.ru}
         \date{\today}
         \begin{abstract}

   Superconductivity of some compounds may be explained as resulting from
   Bose-Einstein condensation (BEC) of atomic electron pairs of divalent
   atoms or electron pairs of diatomic molecules made up of univalent atoms.
   "Artificial" superconductors of such types can be tailored using nonstoichiometric
   compounds. Synthesis of "natural" stoichiometric superconductors is a much more
   complicated problem. In these cases, we have two methods of obtaining dilute metals
   in a state intermediate between the metal and the insulator.
   \end{abstract}
   \maketitle
   \bigskip

  Preparation of materials with preset properties appears simplest and easiest to understand in the case
  of nonstoichiometric compounds - nanocomposites \cite{bib1}. This case may be illustrated by arrays
  of nanostructures embedded in voids of zeolites, asbestos, porous glasses, and opals
  (cluster superlattices) \cite{bib2, bib3,bib4}. Another group of similar systems combines matrices whose
  voids can accommodate sublattices of separate atoms only. This group approaches stoichiometric
  compounds. Actually, it is at the edge of valence interactions between the matrix-filler
  sublattices. Take, for instance, $\textrm{Na}_{x}\textrm{WO}_{3},$ which for $x > 0.33$ becomes a well known tungsten
  bronze.

  Our choice of the $\textrm{Mg}_{x}\textrm{WO}_{3}$
  nanocomposite
  $(x < 0.4)$ may serve as an illustration of a simple
  physico-chemical tailoring of superconductors.
  The filler in the $\textrm{WO}_{3}$ matrix was
  Mg $(2r_{\textrm{Mg}} \sim 3.3$\!\AA; \quad
  the electron pair is formed by two \textit{s} electrons.
  $\Delta_{c} \sim (7.6/20)$\!eV, $T_{cmax} \sim 600$\!K
  for $x \sim 0.4)$ \cite{bib1}.

  The Mg metal in the $\textrm{Mg}_{x}\textrm{WO}_{3}$ system is diluted by insulator in accordance
  with the value of \textit{x} \cite{bib5}. Figure 1\textit{a} shows
  schematically the $\textrm{WO}_{3}$ cell with
  the parameters $a \sim b \sim c \sim 3.78$\!\AA. The void diameter is $\sim 3.8$\AA\, (Fig.\,1c,\,d).
  The Mg atom diameter $d_{\textrm{Mg}} \sim 3.3$\!\AA. The diameter of the "windows" is $\sim 2.7$\!\AA (Fig.\,1b).
  Therefore, penetration of Mg atoms through the windows meets with difficulties, which
  makes $\textrm{Mg}_{x}\textrm{WO}_{3}$ stable. The $\textrm{Mg-WO}_{3}$ contact is close
  to valence interactions, because
  the metal-insulator phase gap is usually $\sim (0.5--0.7)$\!\AA\, \cite{bib6}.

  Figure 2 plots the temperature dependence of the resistance, \textit{R(T)}, of a sample of
  pressed powder $(\textrm{MgO}, \textrm{WO}_{2}, \textrm{WO}_{3})$ containing several
  "perovskite type" $\textrm{Mg}_{(0.037--0.125)}\textrm{WO}_{3}$
  phases. (The sample was prepared by A. V. Golubkov.) One immediately sees several transitions
  in the $100--300$\!K range. The Meissner effect was not observed because of the presence of
  nonsuperconducting phases.

  $\textrm{Mg}_{x}\textrm{WO}_{3}$ is a nanocomposite, and it differs from $\textrm{MgB}_{2},$
  in which Mg is diluted "chemically"
  but which still remains a metal with BCS superconductivity.

  The cubic void array of $\textrm{WO}_{3}$ can accommodate several regular and equilibrium sublattices
  of Mg atoms with a cell volume $V_{1}$ and the number \textit{x} of Mg atoms per
  $\textrm{WO}_{3}$ cell $(\sim a^{3})$Fig.\,2.
   In the case of BEC, to each value of \textit{x} corresponds $T_{cx} \sim x^{2/3} \sim n_{2}^{2/3}.$
   For $x = 0.125\; (\textrm{V}_{1}=2^{3}a^{3}),$
   the electron pair concentration is $n_{2} = 23\cdot10^{20} \textrm{cm}^{-3},$ and $T_{cx} \sim 290$\!K\!\!
   $(m^{*} \sim 10 m_{e}$ \cite{bib1}).
   The Table lists also experimental values of $T_{cexp},$ which are close to $T_{cx}$(Figs.\,2 and 3).

\begin{table}[h]

\begin{tabular}{|c|c|c|c|c|c||c|c|} 
\hline $\textrm{V}_{1}$ & $a^{3}$ & $2(a)^{3}$ & $(2a)^{3}$ & $2^{2}3(a)^{3}$ & $2(2a)^{3}$ & $(3a)^{3}$& $2(3a)^{3}$  \\
\hline $x$ & $1$ & $0.5$ & $0.125$ & $0.083$ & $0.063$ & $0.037$ & $0.019$ \\ \hline $T_{cx}$\!K & ? & ? &
$290$ & $225$ & $185$ & $130$ & $76 $\\ \hline $ T_{cexp}$\!K &----- & -----& $280$ & $240$ & $175$ & $140$ & $<78$?\\
\hline

\end{tabular}
\end{table}

  A condensate of Mg atoms which practically do not interact with the matrix may serve as a model
  for a system with Bose--Einstein electron pair condensation \cite{bib5}.
    Matrices similar to $\textrm{WO}_{3}$ and corresponding standard technologies for perovskites and spinels
     may be used to accommodate other electron-pair atoms to tailor new "artificial" superconductors
     (for instance, a spinel $\textrm{MgAl}_{2}\textrm{O}_{4}).$
     The first attempt at synthesizing an "artificial" superconductor
     was undertaken by Ogg in 1946 $(\textrm{NaNH}_{3})$ \cite{bib7}.
  Tailoring of stoichiometric ("natural") superconductors is an issue of formidable complexity.
  It will possibly require invoking computer simulation \cite{bib8}.
  Development of "artificial" and "natural" superconductors may be considered as essentially
  two methods of obtaining dilute metals in a state intermediate between the metal and insulator \cite{bib1,bib9,bib10}.

\begin{figure*}[tbp]
        \includegraphics[scale=0.8]{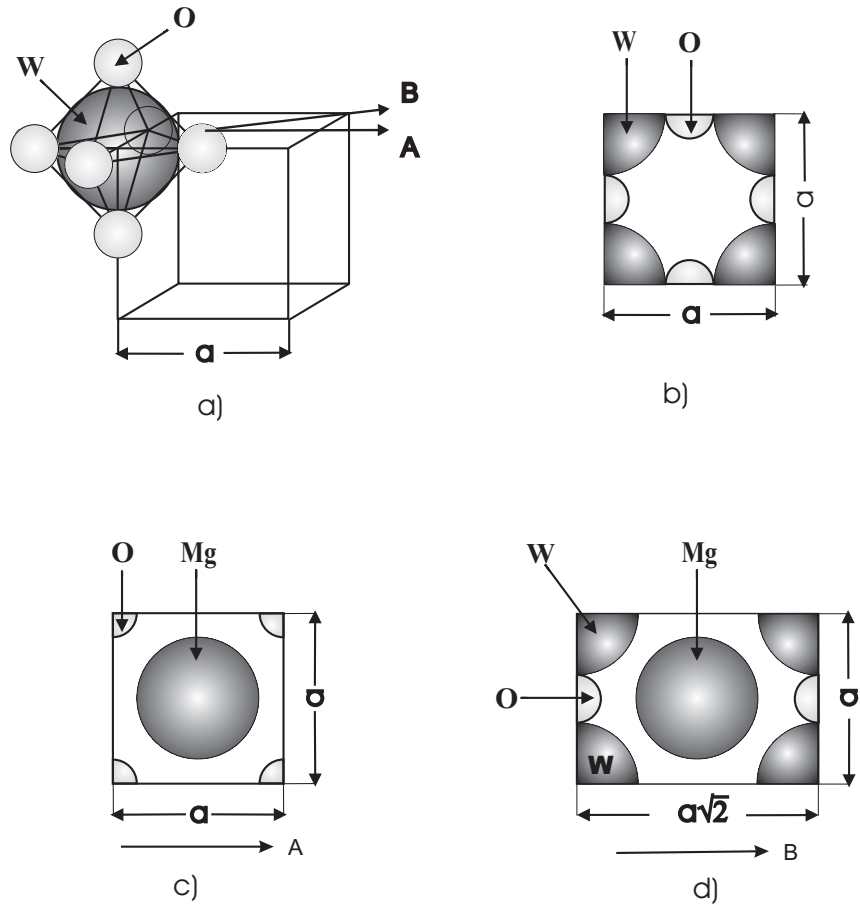}
        \caption{$\textrm{WO}_{3}$ cell (\textit{a});\\
         Side face of  $\textrm{WO}_{3}$ cell (b);\\
         Section of $\textrm{Mg-WO}_{3}$ cell along the A direction (c);\\
         Section of  $\textrm{Mg-WO}_{3}$ cell along the B direction (d);}
        \end{figure*}

\begin{figure*}[tbp]
        \includegraphics[scale=0.8]{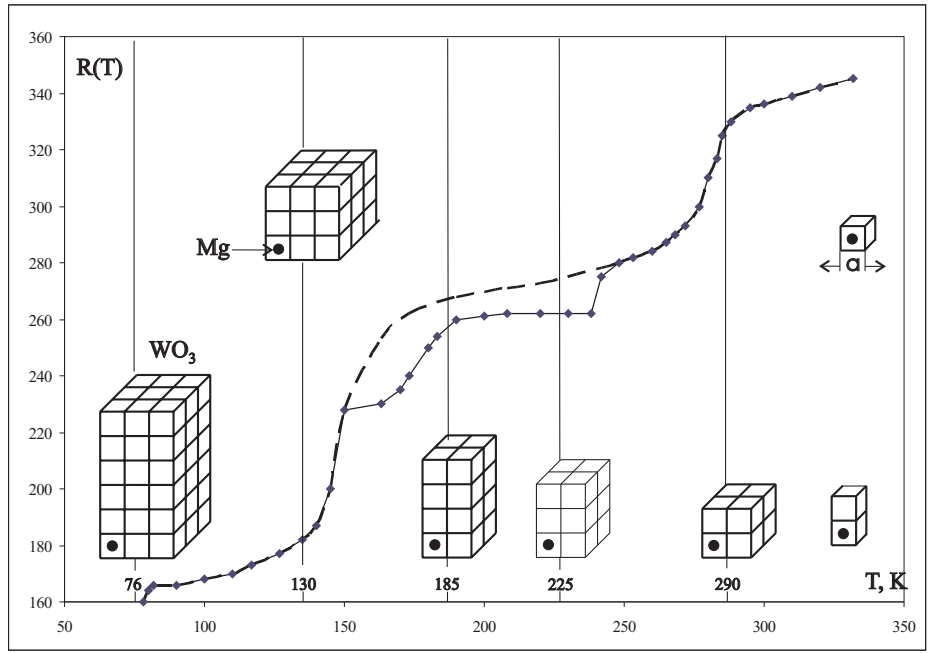}
        \caption{Temperature dependence of the resistance of a $\textrm{Mg}_{x}\textrm{WO}_{3}$ sample.}
        \end{figure*}

\begin{figure*}[tbp]
        \includegraphics[scale=0.8]{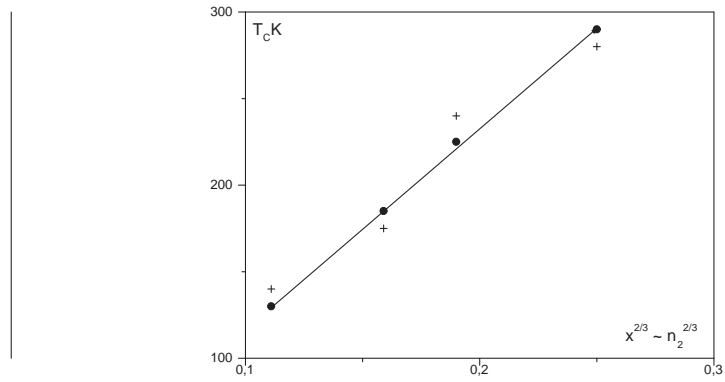}
        \caption{Dependence of $T_{c}$ on $x^{2/3} \sim n_{2}^{2/3}.$
        The same dependence is observed in $\textrm{SrNb}_{x}\textrm{Ti}_{1 -- x}\textrm{O}_{3}$\! \cite{bib5}.\\
        $T_{cx}$ - point; $T_{cexp}$ - cross.}
        \end{figure*}


\begin{thebibliography}{10}

\bibitem{bib1} V.N.Bogomolov,   e-print: http://xxx.lanl.gov/abs/cond-mat/0411574;  /0412740;;  /0505718; /0506399;
/0604509.
\bibitem{bib2} V.N.Bogomolov,   Sov. Phys. Usp.  \textbf{21}, 77  (1978).
\bibitem{bib3} V.N.Bogomolov,  T.M.Pavlova,
Physics and Technics of Semicnnd., \textbf{29}, 826 (1995).
\bibitem{bib4} R.Arita, T.Miyake, T.Kotan, M.van Schilfgaarde,
T.Oka, K.Kuroki, Y.Nozue, H.Aoki, e-print: http://xxx.lanl.gov/abs/cond-mat/0304322.
\bibitem{bib5} V.N.Bogomolov,
e-print: http://xxx.lanl.gov/abs/cond-mat/0406564 ;  /0604509.
\bibitem{bib6} V.N.Bogomolov,   Phys. Rev. B, \textbf{51}, 17040
(1995); Zeolite News Letters, \textbf{10}, 148 (1993).
\bibitem{bib7} P.P.Edvards, J. of Superconductivity: Incorporating Novel Magnetism \textbf{13}, 933 (2000).
(See also R.A. Ogg, Jr.,    Phys. Rev.  \textbf{69}, 243 (1946)).
\bibitem{bib8} A.N.Kolmogorov, S. Curtarolo, e-print: http://xxx.lanl.gov/abs/cond-mat/0603304.
\bibitem{bib9} V.N.Bogomolov, e-print: http://xxx.lanl.gov/abs/cond-mat/9912034;
Techn. Phys. Letters, \textbf{28}, 211 (2002).
\bibitem{bib10} M.Capone, M.Fabrizio, E.Tossatti, e-print: http://xxx.lanl.gov/abs/cond-mat/0101402.
\end{thebibliography}
\end{document}